# Double electric layer influence on dynamic of EUV radiation from plasma of high-current pulse diode in tin vapor

I.V.Borgun[1), N.A.Azarenkov[1], A.Hassanein[2], A.F.Tseluyko[1], V.I.Maslov[1], D.L.Ryabchikov[1]

[1)]*Karazin Kharkiv National University, pl. Svobody 4, Kharkov 61077, Ukraine*
[2)] *School of Nuclear Engineering, Purdue University, 400 Central Drive, West Lafayette, IN 47907-2017, USA*

Generation of high-power pulses of extreme ultraviolet radiation from multi-charged tin plasma are of great interest to many applications. We studied the electric potential distribution in the discharge gap in kind formed double layer. The local plasma heating by electron beam, formed in the double layer, leads to generation of peak radiation pulses and as a result to an increase in the radiation power.

## I. INTRODUCTION

This article is devoted to the generation mechanism of the power (>100 кW) of the short extreme ultraviolet (EUV) pulses in the range of 12.2-15.8nm wavelengths by the multi-charged tin plasma. This is, in particular, concerned with the development of high-power EUV sources for the next generation of nano-lithography.

At the required radiation power ~ 500W for the high volume manufacture, the input power of the advanced radiation sources is about ~300kW [1]. Such power levels make actual search for new ways to increase the conversion efficiency of the input power into the EUV radiation power. On the other hand, for use of high-contrast nonlinear photoresists for nano-lithography the above-threshold pulse intensities are required [2]. This requires the development of new methods for the generation of the power EUV pulses.

Results are presented of a new developed source on the intense EUV pulse generation in the 12.2-15.8 nm wavelength range by plasma diode [3-5]. This work is aimed to investigation of the processes leading to the generation of EUV power pulses in high-current tin-vapor discharges.

## 2. EXPERIMENTS

A schematic of the experimental setup to study the EUV yield from the plasma of multi-charged tin-ions consists of pulsed high-current plasma diode, a scheme for measurement of the radiation intensity in 12.2-15.8nm wavelength range, based on the semiconductor detector AXUV-20, and 3 probes for plasma electric potential measurement. The current in the diode is excited between the rod and tubular electrodes by discharge of the capacitor $C_0$ of 2.0 µF at starting pressure $2\times10^{-6}$ Torr (Fig.3a) after filling discharge space by preliminary plasma. The side surface of a rod electrode is enveloped by tubular ceramic insulator to provide a current density of about 0.1–0.7 MA/cm$^2$.

The diameter of the tubular electrode equals 10 mm, diameter of the rod electrode – 5 mm, and the length of the discharge space is 4.5 cm. The working surface of the electrodes is covered with 0.5-mm-thick tin layer. The discharge voltage is from 4 to 15 kV, the current amplitude – from 5 to 35 kA, half-period of current oscillations – 1.7 µsec.

For the protection of the detector AXUV-20 from plasma and charged particle beams, the radiation is carried out through the tubular channel placed in the transverse magnetic field of 2 kOe. The tubular channel length is 25 cm. To exclude the effect of photoelectrons on the detector signal, a set of shading diaphragms is used and the detector potential is biased by +20 V with respect to the channel.

The probes are set along discharge axis near the rod electrode (Fig. 1). Probe signals are measured using capacitive voltage dividers. An effective probe capacitance is 0.3 pF.

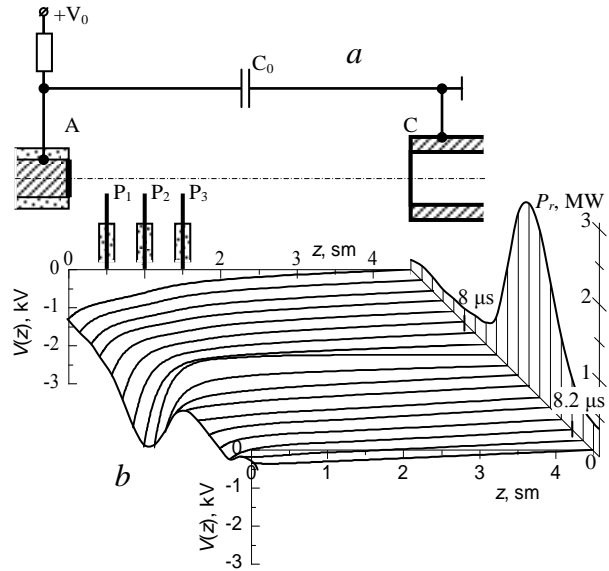

Fig. 1. Schematic illustration of the experimental setup (a) and spatial distribution of electric potential in time (b)

## 3. RESULTS AND DISCUSSION

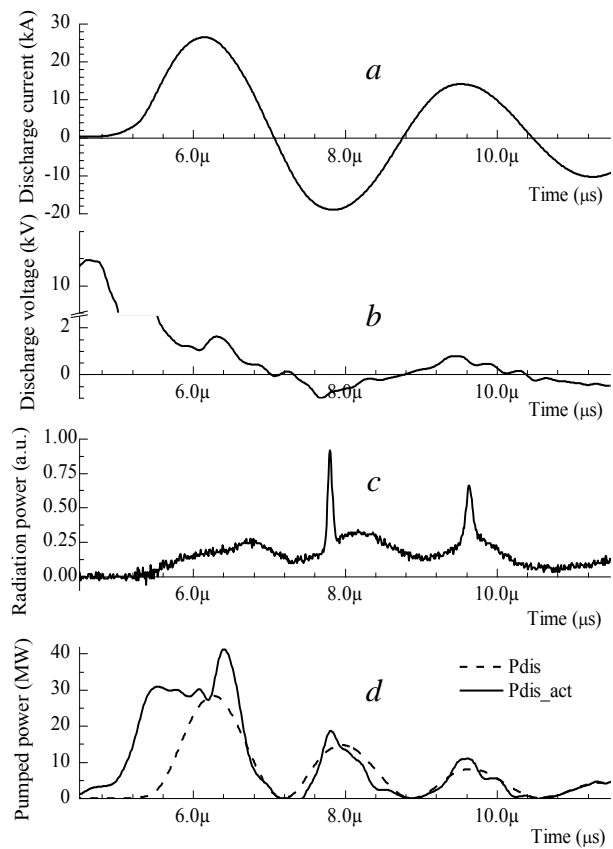

Fig. 2 Temporal dependences of discharge current (a), discharge voltage (b), radiation power (c), and power (d) pumped in the discharge.

Fig. 2 shows the temporal dependences of discharge current (a), discharge voltage (b), EUV intensity in the 12.2-15.8 nm wavelength range (c), and active discharge power (d). It can be seen at the EUV intensity dependence (Fig. 2c) that narrow EUV pulses (spikes) exist in the second and third half-periods of the discharge current oscillation. The EUV spikes are clearly seen against the background of relatively long radiation pulses (main radiation), but are much higher and shorter (~100–200 ns).

Comparison of the curves *c* and *d* shows that the spike radiation pulse is preceded by a sharp increase of the peak input active power. Estimates show that an additional 25% rapid power input leads to an increase of EUV intensity (due to peak pulse) of 50%. In other words, the conversion efficiency of external energy into EUV energy (in case of sudden power input) increased about 2 times.

It should be noted that in contrast to the main radiation, which is observed around the plasma column, the peak radiation comes from a certain local area of the discharge near the rod electrode.

To determine the causes of the peak radiation from the local area the dynamics of the plasma electric potential distribution along the discharge gap is investigated. The results are presented in Fig. 1.

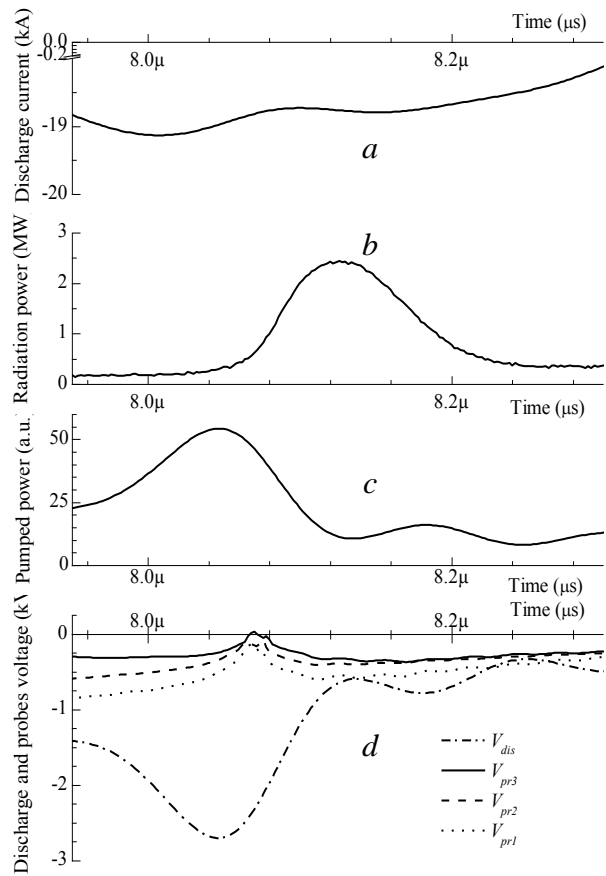

Fig. 3. Temporal dependences of discharge current (a), radiation power (b), pumped power (c) in the discharge, discharge, and probe voltages (d) in the second half-period of the discharge current oscillation

Fragments of the discharge current, voltage, and the potentials of the probes, the radiation intensity, and power inputted into the discharge in the second half-period oscillations of the discharge current are presented. The figure shows that an increase in discharge voltage corresponds to the time of a sharp rise in input power. At the same time the potentials of probes, located at the surface of the rod electrode, decreases. This means that all voltage, applied to the discharge, is concentrated in a narrow region between the electrodes and probes.

Fig. 1 also shows the evolution of the electric potential distribution along the axis of the discharge gap at the time of spike radiation formation. The concentration of the applied voltage in a narrow region demonstrates in favor of the electric double layer formation. Estimation of this double layer thickness, based on the voltage and the discharge current and the electrode area, gives 20 μm. In this case, the mechanism of additional power input is as follows: counter-accelerated electron and ion beams are formed in the electric double layer (the main energy transfers to the electron beam), the electron beam collectively interacts with plasma, adjacent to the rod electrode side of the double layer, and rapidly heats it up. This leads to a sharp increase in EUV radiation from the local area.

## 4. CONCLUSIONS

The use of the spike radiation allows more efficiently to convert input energy into EUV energy and significantly increases the conversion efficiency. At the same time the power of these pulses is quite sufficient to work with nonlinear photoresists. Needed to generate the spike EUV radiation the sharp power input in the discharge is provided due to the formation of the double electric layer. The electron beam formed in the double layer due to the collective interaction rapidly heats plasma electrons to a temperature sufficient to generate EUV radiation. The condition for the double layer formation is the excess of the current, provided by power supply, over current that can be carried by plasma [6]. At high plasma density the area of beam-plasma heating has a small size, that determines the small region of the EUV generation. In the case of electrodes with different working surfaces, the double layer is always formed near the small work surface electrode. Small size and sufficient spatial stabilization of the spike EUV generation region are additional promising reasons for use of the spikes in future nano-lithography.